\documentclass[english,aps,prstper,reprint,showpacs,titlepage,longbibliography]{revtex4-2}   
\usepackage[T1]{fontenc}	
\usepackage{geometry}		
\usepackage{graphicx}
\usepackage[above,below]{placeins}	
\usepackage{times}

\usepackage{hyperref}  
\hypersetup{colorlinks=true,urlcolor=blue,citecolor=blue,linkcolor=blue}   
\usepackage{enumitem}          
\setlist{nosep}                 

\begin{document}

\begin{titlepage}

\title{Unveiling Gender Dynamics in Introductory Physics Labs}

 \author{Bilas Paul}
\email[Please address correspondence to ]{palb@farmingdale.edu}
\affiliation{SUNY Farmingdale State College, Farmingdale, NY, 11735}

\begin{abstract}
\label{abstract}
\noindent 
The persistent underrepresentation of women and gender minorities within the physical sciences remains a significant issue. This study investigates gender dynamics in introductory algebra-based physics laboratories, focusing on participation, task preferences, and comfort levels. Statistical analysis revealed no significant gender difference in overall participation rates during lab activities. However, significant gender-based disparities emerged in both task preference  [\(\chi^2(\text{df}=3) = 9.548,~ p = 0.023, ~ \alpha = 0.05 \)] and comfort levels [\(\chi^2(\text{df}=3) = 7.906,~ p = 0.048, ~ \alpha = 0.05\)]. Male students significantly preferred and felt more comfortable with hands-on equipment handling and data collection, whereas female students more frequently preferred and reported higher comfort with analytical and documentation tasks like note-taking, calculations, and report writing. Qualitative responses highlighted additional challenges reported by some women, including exclusion from group discussions and reluctance to contribute ideas in male-dominated groups. These findings suggest that while overall participation may appear gender-neutral, gendered patterns in task allocation and comfort persist. The results underscore the need for instructional strategies that promote equitable engagement and foster inclusive laboratory environments in physics education.
\end{abstract}

\maketitle
\end{titlepage}

\section{Introduction }
\label{intro}
Despite decades of advocacy and policy initiatives aimed at fostering diversity and inclusion, women continue to be markedly underrepresented in physics and other STEM (science, technology, engineering, and mathematics) disciplines, particularly within academic and research environments. As of 2024, women comprise only about 28\% of the global STEM workforce, with even lower representation in fields such as physics, engineering, and computer science~\cite{swe2025globalstem, aiprm2025womeninstem}.  This persistent gender gap in STEM is  influenced by a complex interplay of factors, including entrenched societal norms, pervasive stereotypes, and systemic institutional barriers that can subtly or overtly shape students' educational trajectories. Introductory physics courses, which are often gateways to numerous STEM majors, serve as a crucial environment in which these barriers can either be reinforced or mitigated. Consequently, these courses represent a vital locus of research for understanding the mechanisms that influence gender equity in STEM.

Recent research emphasizes  that classroom culture and peer dynamics across STEM disciplines are central to understanding the persistent underrepresentation of women. 
 Studies consistently reveal that women tend to participate less frequently in class discussions and laboratory activities compared to their male peers, a pattern that is often reinforced by subtle biases in instructor-student and peer interactions. For instance, women are more likely to be interrupted, less likely to be called upon, and often report feeling overlooked or undervalued during group work. These dynamics can erode self-confidence and contribute to a diminished sense of belonging, both of which are critical for persistence in STEM fields~\cite{aguillon2020genderAL, PhysRevPhysEducRes.17.020140, CASPI2008718, PhysRevPhysEducRes.18.020123}. Furthermore, the structure of classroom activities and assessment formats—such as the prevalence of timed exams or competitive environments—can exacerbate performance gaps by amplifying test anxiety and stereotype threat among female students~\cite{eddy2014gendergaps, maries2024challenges, burkholder2022preinstruction}.

Central to understanding these disparities is the concept of self-efficacy, or students’ beliefs in their own abilities to succeed in STEM disciplines. Recent studies have shown that women in introductory STEM courses—including, but not limited to, physics—consistently report lower self-efficacy than men, even when their academic preparation and performance are comparable.  This gap persists regardless of the numerical representation of women in the classroom, indicating that simply increasing female enrollment is insufficient to address underlying confidence and identity issues. Lower self-efficacy among women is closely linked to reduced participation, higher levels of anxiety, and a greater likelihood of leaving STEM pathways altogether. Factors contributing to this gap include persistent societal stereotypes about who “belongs” in STEM, a lack of visible female role models, and classroom climates that may inadvertently reinforce traditional gender roles~\cite{brage2015gendergap, PhysRevPhysEducRes.18.020142}. 
 
Laboratory and collaborative learning settings, which are integral components of many gateway STEM courses, offer a unique lens through which to examine gendered patterns of engagement and task allocation. 
While some recent multi-institutional studies have found no significant gender-based differences in task preferences and comfort levels, institution-specific patterns in participation did emerge~\cite{paul2025breaking}. Qualitative feedback from that study further revealed persistent challenges among female students, including perceived assumptions about competence, marginalization within group work, and reluctance to voice opinions in male-dominated settings. Other research similarly highlights that women are often relegated to organizational or supportive roles and express hesitancy to assert ideas in group settings, even when overall participation appears comparable~\cite{PhysRevPhysEducRes.20.010102,PhysRevPhysEducRes.16.010129, PhysRevPhysEducRes.20.010121}. These findings underscore the importance of analyzing not only participation rates, but also the distribution of roles, comfort levels, and lived experiences within collaborative learning environments.

Building on our previous multi-institutional analysis of gendered participation in physics labs~\cite{paul2025breaking}, this study employs a focused, in-depth approach to examine gender dynamics within a single institutional context. While our prior research identified broad patterns across three institutions, the data from each setting were analyzed separately, which limited the statistical power of our findings due to the small sample size at each site. The present study overcomes this limitation by analyzing a substantially larger dataset drawn from five course sections at a single institution. This expanded sample enables a more rigorous and detailed statistical analysis than was previously possible, allowing us to move beyond broad comparisons to investigate the nuanced performance, learning preferences, and specific barriers encountered by non-physics STEM majors, with a particular emphasis on female students. The present analysis leverages a consistent institutional environment to control for variables such as curriculum, instructor expectations, and institutional culture, thereby isolating factors that may uniquely influence gendered experiences within this context. 

\section{Materials \& Methods}
This study employed a mixed-methods research design to conduct a multifaceted examination of gender-related patterns in introductory physics laboratories, including student participation, task preferences, and comfort levels.
By integrating both quantitative and qualitative data, the study aimed to provide a comprehensive perspective on how students of different genders interact with laboratory tasks, express preferences, and encounter barriers in collaborative learning environments.

The research was conducted at Farmingdale State College in Farmingdale, New York, a public institution that is part of the State University of New York (SUNY) system. The study focused on non-physics STEM majors enrolled in algebra-based introductory physics courses across five course sections at this institution. A total of 104 students participated, including 64 who identified as male, 39 as female, and one as non-binary. Due to the limited representation of non-binary students, only male and female participants (\(n=103\)) were included in the statistical analyses to ensure meaningful group comparisons. The participant pool reflected the demographic profile of the institution's student body.
According to institutional data, approximately 44.1\% of students identified as White, 27.2\% as Hispanic/Latino, 12.1\% as Asian, and 10.4\% as Black or African American. The gender distribution was roughly 59.0\% male and 41.0\% female, with over half of the students belonging to minority groups. The gender distribution of our analytical sample (64 male, 62.1\% ; 39 female, 37.9\%) was reasonably consistent with that of the overall student body at the institution. The majority of participants were pursuing degrees in fields such as aviation, biology, engineering, engineering technology, and health science, providing a representative cross-section of non-physics STEM majors.

All research procedures were reviewed and approved by the institutional review board at SUNY Farmingdale State College, ensuring compliance with ethical guidelines for research involving human participants. Informed consent was obtained from all participants, who were assured of the confidentiality and anonymity of their responses. Data were collected and stored securely, and participants were informed of their right to withdraw from the study at any time without penalty.

The study focused on Physics II laboratory courses, as students enrolled in these sections had already completed the prerequisite introductory course. This selection ensured that participants had prior exposure to physics concepts and laboratory equipment, thereby reducing the influence of initial unfamiliarity and allowing for a more reliable assessment of engagement and participation patterns among students with foundational knowledge.
During laboratory sessions, students worked in small groups of three or four to complete weekly experiments. The collaborative nature of these sessions created a natural context for observing group dynamics, task allocation, and individual engagement.

To achieve a comprehensive perspective on students’ engagement and attitudes toward laboratory activities, the research methodology integrated both systematic observation and survey-based data collection. A structured observational protocol was implemented to systematically capture student engagement during laboratory activities.  The primary instructor conducted real-time, in-person observations using a cyclic, momentary-time-sampling approach. Each lab section included five to six groups, and the observer rotated among them in a fixed sequence. Observations of each group were intentionally brief—typically lasting 1–2 minutes—before the instructor moved on to the next group. This procedure allowed all groups to be observed approximately every 18–20 minutes without missing engagement opportunities in other groups. Each lab session lasted approximately two hours, providing multiple observation cycles per group.  While this method does not capture every moment of interaction, it provides a representative sample of engagement across all groups and tasks over the duration of the lab. The brief nature of observation also helped ensure that the instructor’s presence remained minimally intrusive, allowing students to work naturally. The decision to use in-person observation instead of video recording was made to comply with institutional review board protocols concerning student privacy and to avoid the potential for increased anxiety or altered behavior among participants that a permanent recording might cause. For each observation, the instructor recorded whether individual students were actively participating in specific task categories, such as setting up laboratory equipment, collecting experimental data, taking notes, performing calculations, plotting graphs, and engaging in group discussions related to the experiment. Data were collected for three representative experiments—Gas Laws, Heat Engine, and Mapping Electric Potential—chosen for their alignment with key Physics II concepts and their capacity to foster active student involvement over the course of nearly two-hour sessions. For each student, the proportion of observation points at which they were engaged in any of the defined activity categories was averaged across the three experiments to create an overall participation percentage. These percentages were aggregated by gender for subsequent comparative analysis. Participation distributions were visualized with histograms, and independent-samples \(t-\)tests were used to assess whether group differences were statistically significant.

To provide clearer context for the laboratory environment, we describe one of the observed experiments, Behavior of a Gas (Gas Laws), which investigates the Ideal Gas Law (\(PV=nRT\)). Students work in small groups to explore three fundamental  relationships:
\begin{itemize}
\item[1.] \textbf{Pressure vs. Volume:} Using a gas syringe and pressure sensor, students varied the system's volume at approximately constant temperature to verify Boyle's Law (\(P\propto 1/V\)).
\item[2.] \textbf{Pressure vs. Number of Particles:}  Students added successive ``puffs'' of air to a syringe at constant volume and temperature to investigate the direct relationship between pressure and the amount of gas.
\item[3.] \textbf{Pressure vs. Temperature:} Using a sealed flask in a water bath equipped with a temperature probe, students heated and cooled the system at constant volume to explore Gay-Lussac's Law (\(P\propto T\)),  extrapolating their data to conceptualize absolute zero.
\end{itemize}

This experiment necessitates a division of labor, creating natural opportunities for role specialization. Key tasks include hands-on equipment handling (assembling apparatus, manipulating syringes, managing hot water), digital data collection (operating the sensor interface, recording measurements), and analytical work (performing curve fits, documenting results, synthesizing relationships). The collaborative and multi-faceted nature of this lab makes it a robust setting for observing gendered patterns in task preference and comfort.

In addition to observational data, a survey was administered to all participants at the conclusion of the laboratory sequence to further elucidate students’ perspectives and preferences regarding laboratory work. The survey included both closed- and open-ended items.  Students were asked to self-identify their gender and to indicate their preferred and most comfortable laboratory activities. Preferences were captured by asking participants to rank four core tasks—(A) equipment setup, (B) data collection, (C) note-taking/calculations/data visualization, and (D) report writing—from 1 (most preferred) to 4 (least preferred).
For comfortability, students rated each laboratory activity on a scale from 1 (least comfortable) to 5 (most comfortable). The survey design allowed students the flexibility to rate multiple activities as 5 if they felt equally comfortable with more than one task, or to refrain from assigning a 5 to any activity if they did not feel particularly comfortable with any of the options. Note-taking, calculations, and data visualization were grouped as a single category, reflecting their frequent co-occurrence during data analysis and their shared cognitive demands~\cite{padilla2018decision, doi:10.1177/1473871611433713} .  To examine possible links between gender and both task preferences and comfort ratings, chi-square tests were conducted, providing a statistical basis for evaluating whether distributions differed significantly by gender.  All statistical tests (independent samples \(t-\)tests and chi-square tests) were conducted at the 95\% confidence level (\(\alpha = 0.05\)).

Alongside the quantitative measures, the survey included open-ended questions designed to capture students’ reflections on their laboratory experiences. One prompt asked participants to describe the factors that contributed to their comfort level with specific laboratory tasks, while others invited them to share any challenges or barriers they encountered during laboratory work, with particular attention to issues related to gender. The qualitative data from these open-ended responses were analyzed using thematic analysis, with an inductive coding approach to identify patterns  related to group dynamics, task allocation, factors influencing comfortability, and gender-related barriers encountered in the laboratory setting. The coding process involved iterative reading and categorization of responses to detect both commonalities and differences across participants’ accounts.

Potential limitations of the study include the possibility of observer bias during data collection and the reliance on self-reported survey data, which may be subject to social desirability effects. Additionally, the exclusion of non-binary participants from statistical analysis, due to small sample size, limits the generalizability of findings to this group.

\section{Results}
\begin{figure*}[htb]
  \includegraphics[width=1\textwidth]{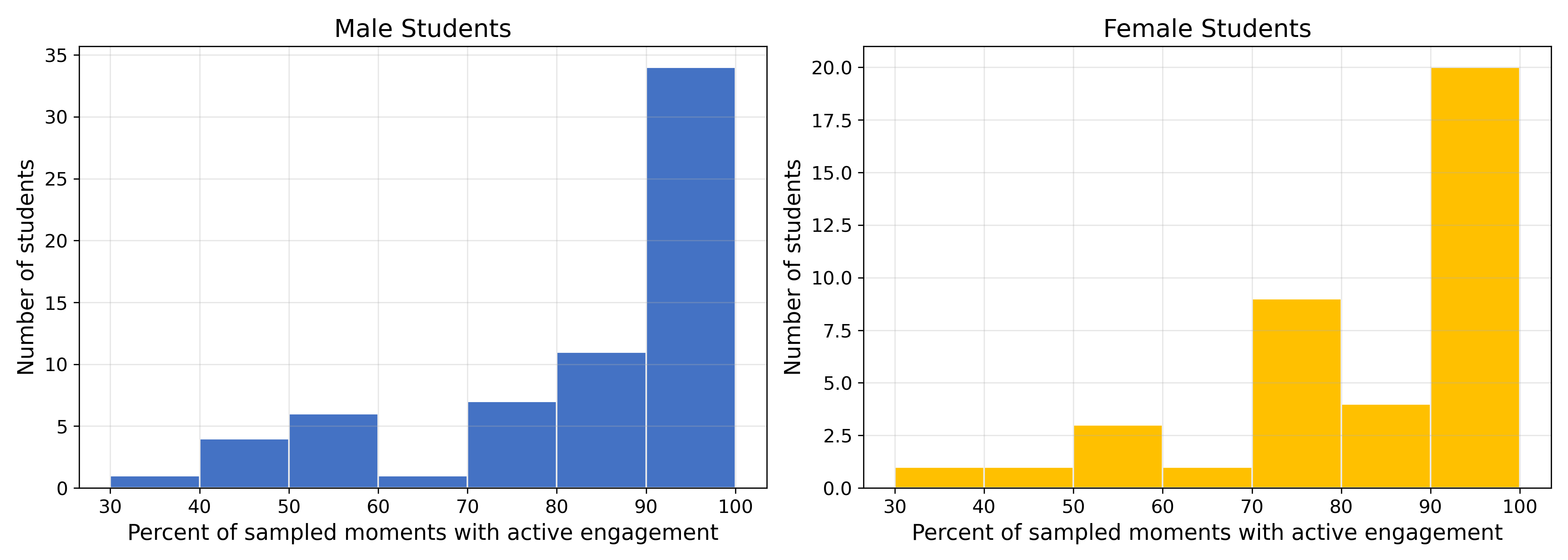}
 \caption{Distribution of students’  participation percentages in physics laboratory sessions. Histograms show the proportion of sampled observations during which individual students were actively engaged in any laboratory activity across three observed experiments.  Blue histogram represents male students (\(n=64\)) while the orange histogram represents female students (\(n=39\)).  \label{fig1}}
\end{figure*}

An analysis of overall participation during laboratory activities (Figure~\ref{fig1}) at State University of New York Farmingdale State College revealed no statistically significant gender differences. Independent samples \(t\)-tests indicated that male and female students engaged comparably in lab activities (\(t= -0.073,~p=0.471, ~\alpha = 0.05\)). This finding is consistent with our earlier small-scale multi-institutional study, in which no significant gender-based differences in participation were observed at SUNY Farmingdale or Valdosta State University, though significant differences were detected at Fairmont State University~\cite{paul2025breaking}. These results suggest that, at least within the context of SUNY Farmingdale, overall participation in laboratory tasks does not differ by gender, regardless of sample size.

\begin{figure}[htb]
  \includegraphics[width=0.48\textwidth]{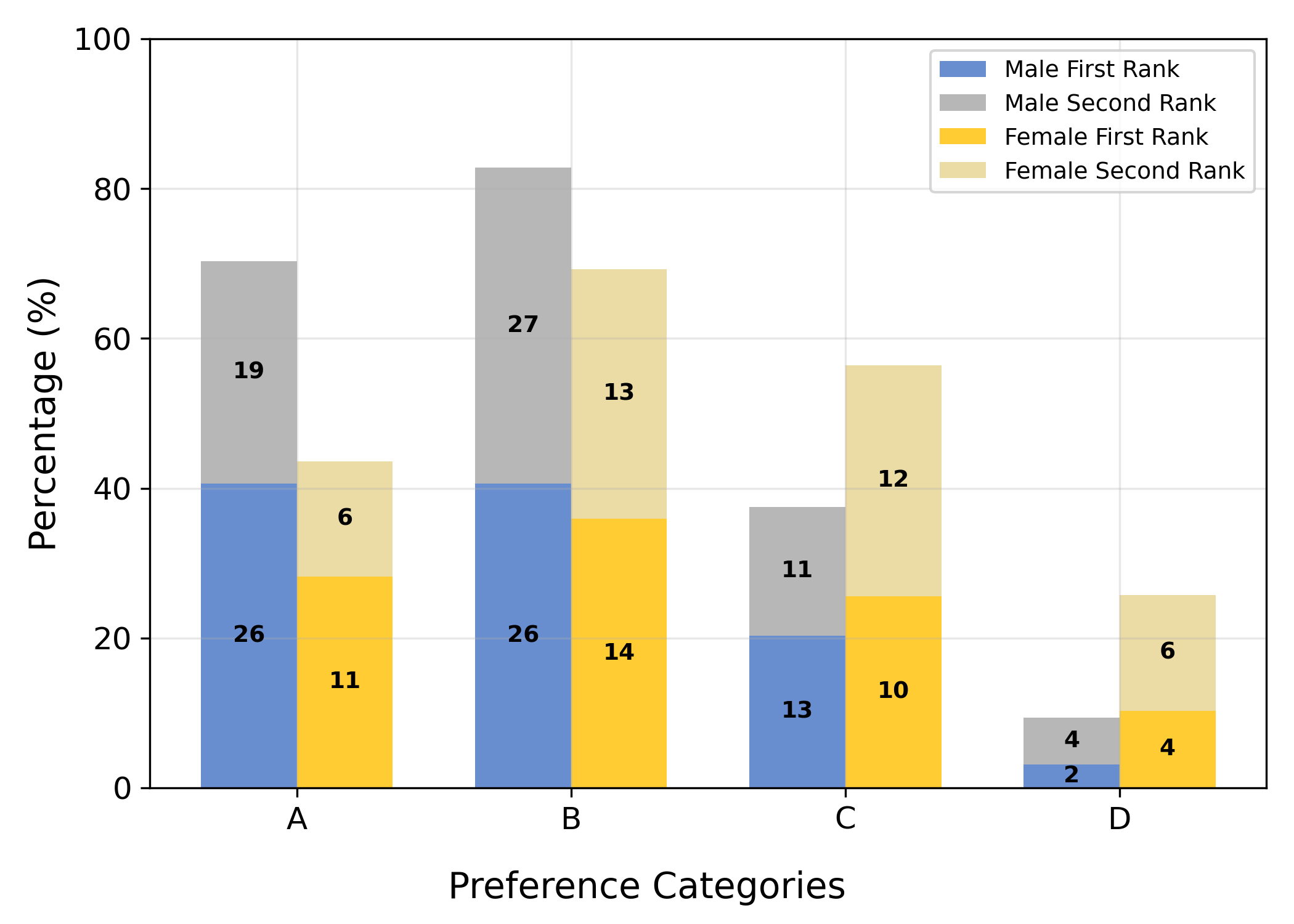}
 \caption{Distribution of students’ preferences for various activities in physics lab sessions. Blue and orange bars represent the preferences of male and female students, respectively, with darker colors indicating first rank and lighter colors representing second rank. Activities: A = Equipment handling, B = Data collection, C = Note-taking, calculations, and plotting graphs, D = Report writing. \label{fig2}}
\end{figure}

To further explore potential gender-based patterns in laboratory engagement, we analyzed students’ stated preferences  (Figure~\ref{fig2}) for specific laboratory tasks, as captured in the post-laboratory survey.  Both first and second ranked task preferences were included in the analysis to provide a more comprehensive view of students’ choices. We conducted a chi-square test of independence to examine the association between gender and task preference, using the aggregated rankings for the four core laboratory activities: equipment setup, data collection, note-taking/calculations/data visualization, and report writing.
The analysis revealed a statistically significant association between gender and laboratory task preference, [\(\chi^2(\text{df}=3) = 9.548,~ p = 0.023, ~\alpha = 0.05\)]. Specifically, male students were more likely to rank equipment setup and data collection among their top preferences, while female students more frequently prioritized note-taking/calculations/data visualization and report writing. This pattern suggests a gendered differentiation in preferred modes of laboratory engagement, with male students gravitating toward hands-on and procedural tasks, and female students favoring roles that emphasize documentation and data analysis.

\begin{figure}[htb]
  \includegraphics[width=0.48\textwidth]{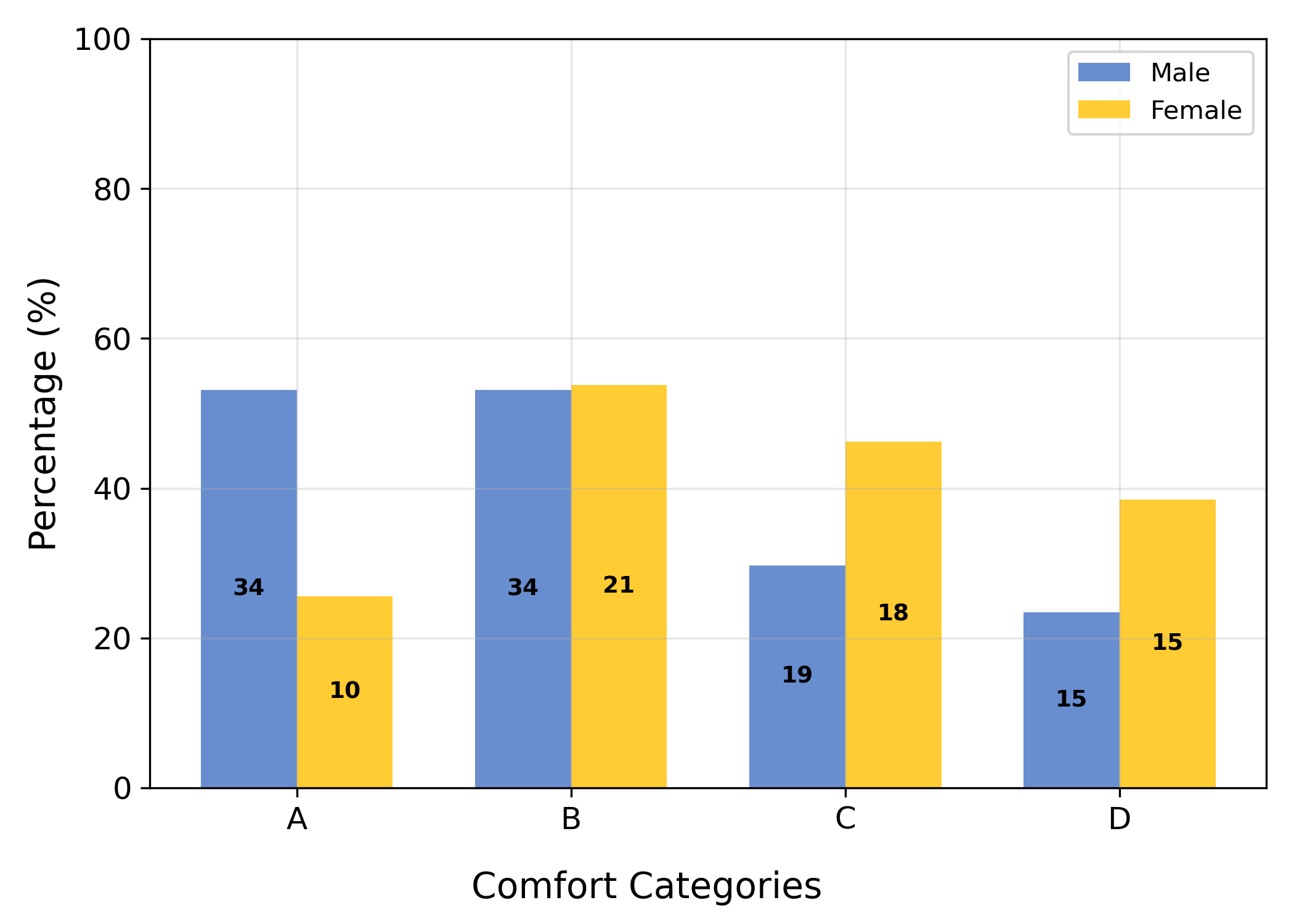}
 \caption{Distribution of students’ comfort levels with various activities in physics lab sessions. Blue and orange bars represent the comfort ratings of male and female students, respectively. Activities: A = Equipment handling, B = Data collection, C = Note-taking, calculations, and plotting graphs, D = Report writing.\label{fig3}}
\end{figure}

To investigate gender-based differences in students’ perceived comfort with laboratory tasks, we analyzed responses in which students identified the activities they felt most comfortable performing  (Figure~\ref{fig3}). For this analysis, a task was classified as a student’s “most comfortable” if it received the highest possible rating (5) on the comfort scale. Notably, students were allowed to assign this top rating to more than one task, resulting in multiple selections per respondent. This approach reflects the multidimensional nature of comfort in laboratory settings, where students may feel equally at ease with several aspects of the experimental process. Conversely, a subset of students  did not report high comfort with any task, as evidenced by ratings below the maximum for all options. These students, who refrained from assigning a 5 to any activity, represent a “low comfort” subgroup within the class. A chi-square test of independence was conducted to assess the association between gender and the selection of most comfortable tasks, accounting for the possibility of multiple selections per student. The analysis revealed a statistically significant relationship between gender and comfort with specific laboratory activities [\(\chi^2(\text{df}=3) = 7.906,~ p = 0.048, ~\alpha = 0.05\)]. Male students were more likely to identify equipment handling as a task with which they felt most comfortable, while female students more frequently selected note-taking/calculations/data visualization  as their top-comfort activities. This pattern closely mirrors the gender-based differences observed in task preferences, suggesting a strong alignment between the roles students prefer and those in which they feel most confident.

Notably, these findings diverge from our previous small-scale multi-institutional study, in which no significant gender associations with laboratory task preferences or comfort levels were observed~\cite{paul2025breaking}. The larger sample size in the present study may have increased the statistical power to detect subtle gender-based differences that were not apparent in the earlier, smaller dataset. This underscores the importance of sample size and context in identifying detailed patterns of participation, preference, and comfort within STEM laboratory environments.

\begin{figure}[htb]
 \includegraphics[width=0.48\textwidth]{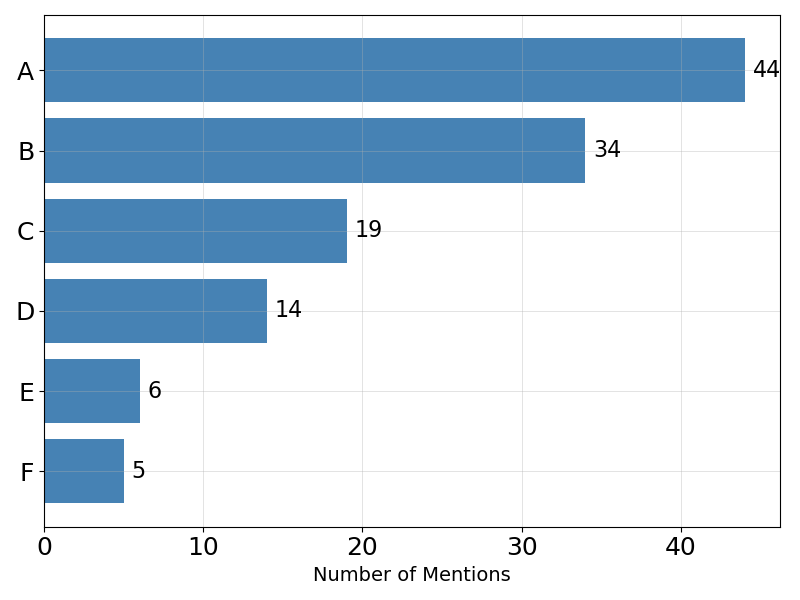} 
\caption{Frequency of major themes identified in students’ open-ended responses about comfort and confidence in physics lab experiments.  The most cited theme was (A) Prior Knowledge and Experience, followed by (B) Group Dynamics and Peer Support. Other significant factors included (C) Instructor Support, (D) Clarity of Instructions, (E) Gender Representation, and (F) Anxiety and Fear of Mistakes. \label{fig4}} 
\end{figure}

To further contextualize these quantitative findings, we conducted a thematic analysis of students’ open-ended responses regarding factors that contributed to their comfort and confidence in conducting physics lab experiments. Thematic analysis revealed several recurring themes, which are summarized below along with their frequency of occurrence (Figure~\ref{fig4}).

\begin{itemize} 

\item \textbf{Prior Knowledge and Experience}: Comfort was often linked to previous experience with laboratory work or familiarity with the course material. For instance, “\textit{My level of confidence/comfort is improved by my studying and how well I know the content.}” 

\item \textbf{Group Dynamics and Peer Support}: Students frequently cited the importance of supportive, collaborative, and reliable lab partners. For example, one student wrote, “\textit{Having a supportive lab group that works together and helps each other out when facing challenges}” was a key factor in their confidence. 

\item \textbf{Instructor Support and Positive Learning Environment}: Many students highlighted the value of accessible and helpful instructors. As one student noted, “\textit{The professor helping me set up the experiment… helps us feel more comfortable and confident conducting physics lab experiments.}”

\item \textbf{Clarity of Instructions and Lab Materials}: Clear, detailed instructions and well-structured lab guides were seen as important contributors to comfort. One student stated, “\textit{The more detailed the lab steps are, the more confident I feel doing it.}” 

\item \textbf{Gender Representation and Stereotype Threat}: Several students, particularly women, described the positive impact of gender representation in their lab groups and the challenges of stereotype threat. For example, “\textit{More than half of my class is made up of women, so I do not feel like there is any ‘impostor syndrome’ or ‘stereotype bias’ affecting my work in the lab.}” 

\item \textbf{Anxiety, Fear of Mistakes, and Feeling Overlooked}: Among those in the low comfort group, common sources of discomfort included anxiety about making mistakes, lack of prior experience, and feeling overlooked or undervalued in group settings. For example, “\textit{... I felt confident w/ material and writing reports/answering questions but felt lost when it comes to using materials.}” \end{itemize}

A demographic analysis of the “low comfort” subgroup revealed that it consisted of 8 female students (3 White, 1 Asian, 2 Hispanic, and 2 Black) and 11 male students (2 White, 3 Asian, 6 Hispanic). This composition suggests that students from underrepresented backgrounds—including women and students of color—are disproportionately represented among those who feel least comfortable in the lab. This pattern is consistent with prior research indicating that women, gender minorities, and students with less prior exposure to laboratory environments often experience heightened discomfort or marginalization in STEM settings~\cite{PhysRevPhysEducRes.20.010102}.
The observed consistency between students’ preferred and most comfortable tasks highlights the interplay between self-perceived competence and role selection in laboratory environments. At the same time, the existence of a “low comfort” subgroup underscores the need for intentional instructional strategies to promote equitable participation and comfort across all laboratory roles, ensuring that all students—not just those who already feel confident—are supported in developing the skills and self-efficacy necessary for success in STEM fields.

To further explore the presence of gender-related barriers in laboratory settings, students were asked whether they had experienced or witnessed any challenges related to gender while conducting physics lab experiments. Out of the total sample, only 15 students (10 female and 5 male) answered  ``yes'' to this question, representing a small subset (approximately \(~15\%\)) of the overall participant pool. Notably, only two female students and no male students chose to elaborate on their experiences in the open-ended response section.

One student described the discomfort of being the only woman in her lab group, which made it difficult to speak up or correct a group member’s mistake due to fear of being discredited or dismissed:

“\textit{Being in a lab group where I am the only girl made it difficult for me to feel comfortable enough to speak out on my opinion. I was often too nervous to try and correct a lab member’s mistake with the fear my lab member may discredit my finding or shrug off my input since I am just a girl. I suggest having at least one pair of like genders in a group. For example, if there are x number of boys, have at least 2 girls in that group. Or maybe form a group with only like genders. I believe the latter to be the best option.}”

Another student highlighted the impact of physical lab arrangements and gendered assumptions on her comfort and participation:

“\textit{I am not very comfortable with setting up the experiment because of the separation at the tables. I feel like my male partners being on opposite side of the tables makes setup difficult because we cannot be all on same side. This makes me feel clueless at the setup stage. Also, sometimes as a woman you just assume that the male partners will know more and you do not want to mess up the steps.}”


These qualitative insights, though drawn from a very small number of respondents, provide concrete examples that contextualize the quantitative findings. The lived experiences described above illustrate potential barriers related to group dynamics, role allocation, and gender representation that can exist in laboratory settings, as established in the broader literature~\cite{Charlesworth2019GenderSTEM, 10.1119/perc.2022.pr.dew, 10.1088/1361-6404/abd597, 10.1088/1361-6404/ab7831, PhysRevPhysEducRes.16.010129, 10.1103/physrevphyseducres.14.020123}. While these specific accounts are from only two students and are not representative of the entire cohort, they offer valuable, context-rich examples of the challenges that can negatively impact the engagement and sense of belonging for some students, particularly women in male-dominated groups. It is also noteworthy that five male students reported witnessing or experiencing gender-related challenges, though none provided detailed written accounts.
The limited qualitative data underscore the need for future research with larger samples and more systematic follow-up to understand the full spectrum of gender-related experiences in physics laboratories.

\section{DISCUSSION}
This study examined gender dynamics in introductory physics laboratory classes at SUNY Farmingdale State College, focusing on gender-related patterns in participation, task preferences, comfort levels, and perceived barriers in lab settings. The results offer important insights into how gender shapes students’ experiences and point to ways to make physics education more equitable.

The quantitative analysis of overall participation patterns in laboratory tasks revealed no statistically significant gender differences (\(t= -0.073,~p=0.471,~\alpha = 0.05\)). This suggests that, in terms of the amount of time male and female students spent on various lab activities, their overall involvement was comparable within this observed context.  This aligns with findings from other studies that emphasize the complexity of identifying participation disparities solely through time-based measures~\cite{ballen2019smaller}.

A significant aspect of this study involved analyzing students' task preferences and comfort levels. In contrast to smaller-scale investigations~\cite{paul2025breaking}, the present study revealed statistically significant differences in how male and female students reported their task preferences [\(\chi^2(\text{df}=3) = 9.548,~ p = 0.023, ~\alpha = 0.05\)] and comfort levels [\(\chi^2(\text{df}=3) = 7.906,~ p = 0.048, ~\alpha = 0.05\)]. Specifically, male students showed a preference for equipment handling and data collection, while female students favored note-taking, calculations, and plotting. Similarly, male students reported higher comfort with equipment handling, while female students reported greater comfort with the more organizational or analytical aspects of lab work.  hese findings indicate that while overall participation time appeared gender-neutral, underlying differences in students' preferred roles and comfort zones can reflect deeper, gendered patterns of engagement. Such differences may be shaped by a range of influences, including prior experience, classroom dynamics, and societal expectations. Importantly, these patterns point to the need for instructors to consider not just how much students participate, but how they participate—and whether certain roles become informally gendered within group settings.

Qualitative analysis of open-ended survey responses highlighted several factors contributing to students’ comfort in the laboratory. Students most often cited prior knowledge and experience, supportive group dynamics, helpful instructors, and clear instructions as factors that increased their comfort. Conversely, discomfort was associated with unclear instructions, lack of experience, and feeling overlooked or undervalued within group settings. Notably, some female students specifically mentioned the impact of gender representation and stereotype threat, describing hesitancy to participate fully in male-dominated groups and concerns about being judged based on gender. These themes are consistent with prior research emphasizing the importance of social and instructional support, clear communication, and attention to gender dynamics in fostering comfortable and engaging STEM lab environments~\cite{shapiro2012role, PhysRevPhysEducRes.18.010106}.

Although only a minority of students provided detailed accounts of gender-related barriers, the experiences shared by a few female students included feeling assumed to be less competent by male peers, being overlooked during group work, and reluctance to share opinions for fear of being dismissed. It is important to note that several male students also reported gender-related challenges, albeit without detailed elaboration, suggesting that gender dynamics can be complex and affect multiple groups in different ways.
While these experiences were not universal, they reflect well-documented issues such as implicit bias, stereotype threat, and exclusion in STEM learning environments~\cite{Charlesworth2019GenderSTEM, 10.1119/perc.2022.pr.dew, 10.1088/1361-6404/abd597, 10.1088/1361-6404/ab7831, PhysRevPhysEducRes.16.010129, 10.1103/physrevphyseducres.14.020123}. The fact that these issues were reported by a minority of students does not diminish their significance; rather, it underscores how subtle gender dynamics can affect participation and a sense of belonging, even when overall participation rates do not show differences. This highlights the value of qualitative methods in understanding equity in STEM education, as aggregate statistics can obscure the lived experiences of marginalized groups.

Based on the findings of this study and the broader literature, several strategies are recommended to promote gender equity in physics laboratory education~\cite{PhysRevPhysEducRes.20.010102, PhysRevPhysEducRes.16.010129, PhysRevPhysEducRes.18.010102, 10.1119/perc.2020.pr.kalende}. First, structured role rotation should be implemented so that all students gain experience and confidence in each aspect of lab work, which helps prevent informal gendered divisions of tasks. Additionally, intentional group composition is important; forming laboratory groups with attention to gender balance and diversity can reduce feelings of isolation and stereotype threat, especially for women and gender minorities. Active facilitation and support from instructors are also crucial—educators should monitor group dynamics, intervene to redistribute participation when necessary, and provide explicit support and validation for students from underrepresented groups. Furthermore, bias awareness training for faculty and teaching assistants can help them recognize and address implicit biases and stereotype threat in laboratories and classrooms. Finally, fostering an inclusive lab environment by promoting a culture of respect and addressing both physical and social barriers to participation is essential. These recommendations aim to create laboratory environments where all students can participate fully and comfortably, regardless of gender, thereby supporting not only equal participation but also genuine comfort, confidence, and a sense of belonging for all students in STEM labs.

\textbf{Limitations and Future Directions:} This study is limited by its single-institution context and the very small number of detailed qualitative responses, which may not capture the full range of student experiences. Additionally, the exclusion of non-binary students from statistical analysis due to sample size constraints highlights the need for more inclusive research designs in the future. Further research should explore the intersectionality of gender with other identities and investigate the effectiveness of specific interventions for promoting equity in physics laboratory courses.

\acknowledgments{}
The authors extend their gratitude to their students for their participation in this study.

\bibliographystyle{unsrt}
\bibliography{main}

\end{document}